\documentstyle[12pt]{article} 
\begin{document} 
\title{ FCNC and non-standard soft-breaking terms in weak-scale
Supersymmetry
\thanks{Work supported by CONACYT and SNI (M\'exico).}}
\author{ J. Lorenzo Diaz-Cruz
\thanks{email: ldiaz@sirio.ifuap.buap.mx} \\
   Instituto de Fisica, BUAP,  \\
 Ap. Postal J-48, 72500 Puebla, Pue., Mexico \\
      }
\maketitle
\begin{abstract}
We study the inclusion of non-standard soft-breaking terms in 
the minimal SUSY extension of the SM, considering it as a model
of weak-scale SUSY. These terms modify the
higgs-sfermion interaction and the sfermion mass matrices,
which can induce new sources of flavour violation.
Bounds on the new soft parameters can be obtained from current data.
The results are then applied to evaluate
the FCNC top quark decays $t \to c + h_i$ $(h_i=h^0,H^0,A^0)$.
 Implications of complex soft parameters for CP-violation
are also addressed.
\end{abstract}


\newpage

1.- Supersymmetric (SUSY) extensions of the Standard model (SM)
\cite{susyrev} have been extensively studied, mainly because of the
possibilty to solve the hierarchy problem.
The minimal SUSY SM (MSSM) \cite{mssmrev}, has been used as
as a framework to search for signals of SUSY.
The required breaking of SUSY is incorporated in the model
through soft-breaking terms \cite{girardelo}, which include  
gaugino and scalar masses, as well as trilinear interactions.
 General soft-breaking terms can produce large flavour changing neutral
currents (FCNC) \cite{murayama}. Possible solutions to this problem
have been proposed within the main theoretical frameworks of
SUSY-breaking\cite{dinetal}.

The MSSM reproduces the SM agreement with data, and
predicts new signatures associated with
the superpartners that are expected
to appear in current or future colliders
\cite{mssmphen}. However, this anaysis usually involves
some simplifications about the soft-breaking parameters.
For instance, one could work within a particular GUT model and
incorporate some specific mechanism of SUSY breaking, then
use the structure of the soft-terms to
study the mass spectrum of superpartners, evaluate
production cross-section and decay rates,
and search for their signatures at future colliders.
Although this approach makes a certain amount of sense,
one could question its generality and whether the future colliders will
test weak-scale SUSY or only a particular model of SUSY breaking.
 In order to study, in a general setting, the  possible presence
of SUSY in nature, we shall define the MSSM at the weak-scale
by considerig the most general structure of soft-breaking
terms, whose values will be constrained by low-energy
phenomenology.

Although it is widely stated that the soft-terms included in
the definition of the MSSM are the most general ones,
there are extra terms that are not usually considered in
the literature \cite{hallrandall,jackjones}, which 
should  be included in a model-independent analysis of
weak-scale SUSY.
In this paper we study how  the inclusion of non-standard terms
in the MSSM, modify the Higgs-sfermion interactions and
the sfermion mass matrices, which in turn can induce 
new sources of flavour violation.
We evaluate then the contribution of the trilinear terms to the
the FCNC top quark decays $t \to c + h_i$, with $h_i$ denoting the
neutral Higgs bosons of the MSSM.
We also comment on the implication of complex trilinear terms
for CP-violation phenomena.

 2.- The usual trilinear terms included in the MSSM correspond to
interactions of the sfermions with the Higgs doublets ($H_{1,2}$),
of the form
\begin{equation}
 {\cal{L}}_3 = \epsilon _{ij}
  [A^d \tilde{Q^i} H^j_1 \tilde{D}- A^u \tilde{Q^i} H^j_2 \tilde{U} +
  A^l \tilde{L^i} H^j_1 \tilde{E} ] ,
\end{equation}
where $\tilde{Q}, \tilde{L}$ represent the squark and slepton doublets,
whereas the squark and slepton singlets are denoted by
$\tilde{U}, \tilde{D},\tilde{E}$.
Equation (1) resembles the Yukawa Lagrangian of the MSSM,
provided that the fermion fields are replaced by their scalar superpatners.
However, one could write extra soft-breaking terms that resemble
the most general two-Higgs doublet model, known as model III
\cite{thdmIII}, by allowing each sfermion flavour  to
couple to both Higgs doublets, namely,
\begin{equation}
 {\cal{L'}}_3 = \epsilon_{ij} [
         C^d \tilde{Q^i} H^{cj}_2 \tilde{D}-
         C^u \tilde{Q^i}H^{cj}_1 \tilde{U}
      +  C^l \tilde{L^i} H^{cj}_2 \tilde{E} ] ,
\end{equation}
where $H^c_n=i \tau_2 H^*_n$ ($n=1,2$); $A^{u,d,l}$ and
$C^{u,d,l}$ denote $3\times3$ matrices in flavour space.
These terms are indeed soft, because each of the
scalar fields carries $U(1)_Y$ charges that forbidds 
their appearence in tadpoles graphs, which
are the only diagrams that could generate quadratic divergences 
from these cubic interactions \cite{jackjones}.
The resulting squared sfermion mass-matrices ($6\times 6$)
can be written in terms of $3\times 3$ blocks,
as follows
\begin{equation}
M^2_{\tilde f}=  \left(
\begin{array}{ll}
 (M^2_{\tilde{f}})_{LL} &   (M^2_{\tilde{f}})_{LR} \\
 (M^2_{\tilde{f}})^{\dagger}_{LR} &   (M^2_{\tilde{f}})_{RR} 
\end{array}
\right) 
\end{equation}
The mass terms $(M^2_{\tilde{f}})_{LL,RR}$ receive contributions
from the F- and D-terms, after the Higgs fields aquire v.e.v.'s
$<H^0_{1,2}>= v_{1,2}$, as well from the chiral-conserving soft-masses.
On the other hand, the chirality-changing mass terms
$(M^2_{\tilde{f}})_{LR}$, which receive contributions from
F-terms and from the $A-$ and $C-$trilinear interactions,
are given by
\begin{eqnarray}
 (M^2_{\tilde{u}})_{LR} &=& \mu m^0_u \cot \beta+ A^u v \sin \beta +
 C^u v \cos \beta ,  \\
 (M^2_{\tilde{d}})_{LR} &=& \mu m^0_d \tan \beta+ A^d v \cos \beta +
 C^d v \sin \beta ,  \\
 (M^2_{\tilde{l}})_{LR}  &=& \mu m^0_l \tan \beta+ A^u v \cos \beta +
 C^l v \sin \beta ,  
\end{eqnarray}
where $m^0_{u,d,l}$ denote the (non-diagonal) fermionic mass
matrices and $v^2=v^2_1+v^2_2$, $\tan \beta=v_2/v_1$.

 The fermion and sfermion mass matrices must be
diagonalized in order to get the mass eigenstates. However, since
the general fermion and sfermion mass matrices are not diagonalized
by the same rotations, flavour-violating
interactions will appear in the MSSM \cite{fcncrev}.
In our case, since the $C^f$ terms modify the chirality-changing
(LR)  sfermion mass matrices, they can represent a
new source of flavour violation.

3.- To determine
the phenomenological predictions of the model, we need to know
the values of the parameters $A^f$ and $C^f$, which
requires a complete understanding of the mechanism
of SUSY breaking.
In supergravity/superstrings \cite{sugrarev}, these terms
are associated to non-holomorphic interactions, whereas in models with
horizontal symmetries \cite{susyhor},
they will appear as higher-dimensional operators.
In gauge-mediated models \cite{gmmrev},
the non-standard soft-terms will appear as
higher-order loops, as the $A$-terms do (two-loop level).
Thus, the $C^q$ parameters appear to be small in the minimal
realization of the these SUSY-breaking scheemes.
However, their contribution to low-energy processes may not be negligible
when compared with the A-terms, for instance when they are proportional
to the light fermion masses.
Thus, the corresponding $C^{q,l}$ parameters should be included in 
a model independent analysis of FCNC phenomena.

 To discuss FCNC bounds, it
is convenient to work in the so-called super-KM basis,
where fermion mass matrices and fermion-sfermion gaugino
vertices are diagonal; flavour violation arises from the
off-diagonal components of the sfermion mass matrices, which are
treated as mass-insertions in loop-graphs \cite{masieroetal}.
The FCNC bounds on  $M^2_{LR}$ are expressed in terms of
dimensionless parameters:
\begin{equation}
(\delta^{\tilde q}_{LR})_{ij} = \frac{1}{  m^2_{\tilde{q}}  }
         [ V^{q}_L (M^{2}_{\tilde q})_{LR} V^{q\dagger}_R ]_{ij} 
\end{equation}
where $V^q_{L,R}$ denote the diagonalizing matrices of
the fermion masses.
 Bounds on the off-diagonal elements of
$\delta^{\tilde f}_{LR}$ could be obtained,
for instance, by requiring that the SUSY contribution to the 
$K-\bar{K}, D-\bar{D}, B-\bar{B}$ mass differences,
saturates the observed values. Similarly,
the diagonal elements  $(\delta^{\tilde f}_{LR})_{ii}$
can be bounded using the SUSY correction to the
fermion masses.
For d-type squarks, the bounds corresponding to
$m^2_{\tilde{q}}=m^2_{\tilde{g}}=500$ GeV,
are \cite{masieroetal}:
\begin{equation}
(\delta^{\tilde d}_{LR})  \simeq 
\left( 
\begin{array}{lll}
 1.6 \times 10^{-3}    & 4.4 \times 10^{-3} & 3.3 \times 10^{-2} \\
 4.4 \times 10^{-3}   & 2.4 \times 10^{-2}  &  1.6 \times 10^{-2}  \\
 3.3 \times 10^{-2}   & 1.6 \times 10^{-2}  &  7.3 \times 10^{-1}  
 \end{array}
 \right) 
\end{equation}

The C-terms appear in the definition
of the $\delta_{LR}$ parameter, namely:
\begin{equation}
(\delta^{\tilde q}_{LR})_{ij} =
          \frac{1}{  m^2_{\tilde{q}}  }
          ( a_q v \bar{A^q}+ b_q \mu m_q +c_q v \bar{C^q} )
\end{equation}
where  $\bar{A^q}=V^q_L A^q V^{q\dagger}_R$,
$\bar{C^q}=V^q_L C^q V^{q\dagger}_R$;
$m^2_{\tilde{q}}$ denotes an average squark mass,
and $m_q$ is the quark mass matrix;
$a_q, c_q$ can be read from Eqs. (4-6).
However, FCNC data constraints the
off-diagonal elements of the combination
$ A^d \cos \beta + C^d  \sin \beta$ and
$ A^u \sin \beta + C^u  \cos \beta$,
and the constraints are 
strong only for $A^d$ and $C^d$
associated with first and second families.
Moreover, since the analysis of FCNC constraints
is not complete for stop/scharm parameters,
one can only estimate  
$A^u$ and $C^u$ to be in the range $100-1000$ GeV,
for which the $\delta^u_{LR}$ parameters would be
one or two orders of magnitude larger than those of
the  third-family d-type sfermions, still in agreement
with present FCNC bounds.

4.- To illustrate the effects of the non-standard soft-breaking terms,
we shall consider the  FCNC decays of top quark 
$t \to c + h_i$ \cite{tchsup},
including only the contribution arising from
the FCNC Higgs-sfermion interaction,
with the gluino and squarks circulating in the loop.
The resulting expression for the decay width is
\begin{equation}
\Gamma (t \to c +h_i) = \frac{m_t}{16 \pi} (1- \frac{m^2_h}{ m^2_t})
            ( |F_L|^2 + |F_R|^2 ) ,
\end{equation}
where:
\begin{eqnarray}
 F_L &=& \frac{\sqrt{2} \alpha_s}{ 3 \pi} M_{\tilde{g}} r_{h_i}
         C_0( m_{\tilde{t}L},m_{\tilde{g}},m_{\tilde{c}R},
              m^2_t,m^2_c,m^2_h) ,  \\
 F_R &=& \frac{\sqrt{2} \alpha_s}{ 3 \pi} M_{\tilde{g}} r_{h_i}
         C_0( m_{\tilde{t}R},m_{\tilde{g}},m_{\tilde{c}L},
              m^2_t,m^2_c,m^2_h) ,
\end{eqnarray}
$C_0$ denotes the scalar Veltman-Passarino scalar function;
$m_{\tilde t}$, $m_{\tilde c}$, $m_{\tilde{g}}$ correspond to the
stop, scharm and gluino masses, respectively, with
\begin{equation}
\begin{array}{ll}
r_{h_i}= & \left\{
\begin{array}{ll}
 A^u \cos \alpha -   C^u \sin \alpha ,& \, {\rm{for}}  \, h^0 , \\
 A^u \sin \alpha +   C^u \cos \alpha ,& \, {\rm{for}}  \, H^0 , \\
 A^u \cos \beta +   C^u \sin \beta   ,& \, {\rm{for}}  \, A^0 . 
\end{array} 
\right.
\end{array}
\end{equation}
Including only the A-term,
the resulting branching ration has values of order $10^{-5}-10^{-6}$.
On the other hand, if we include $A^u_{tc}$ and $C^u_{tc}$ terms
of similar strenght ($\simeq$ 500 GeV),
we find that the branching ratio reaches values of order $10^{-4}$.
If we also include the constributions from off-diagonal
terms in $M^2_{LL,RR}$ it is possible to obtain branching ratios
of order $10^{-3}$, which could be tested at LHC.
Some representative values of B.R. are shown in table 1.

5.- Another interesting aplication of the new soft-breaking terms
is in CP-violation phenomena. In a recent paper
\cite{masiemura}, it has been proposed to use a non-minimal expression for
the $A$-terms, in order to explain the recently observed value of
$\epsilon'/\epsilon$ as having a SUSY origin. Since the C-terms
can also be complex, its contribution to the imaginary part of
$({\delta^d_{LR}})_{12}$ could enhance the amount of CP-violation
due to SUSY, and would help to explain the observed effect within
the MSSM.

CP-violating Higgs interactions will also receive a contribution from
the $C^f$ terms. For instance, the 
parameter $\eta^l_{CP}$, which measures CP-violation 
in the coupling of Higgs bosons with leptons \cite{cpkoldaetal},
receives a new contributions from the C-terms, with sleptons and
gauginos circulating in the loop, it is given by
\begin{equation}
 \eta^l_{CP}= - \frac{6\alpha_{em}}{20\sqrt{2} \cos^2{\theta_W }y_l}
                 Im[C^l M_1 f(M_1,m_{\tilde l})] ,
\end{equation}
where $y_l$ denotes the Yukawa coupling of lepton $l$,
$m_{\tilde l}, M_1$ corresponds to the slepton and Bino masses, respectively;
$f$ is a function that arises
from the loop integration.
For SUSY masses of order $200$ GeV, $\tan\beta=10$ and $m_A=100$ GeV,
we find that $\eta^{\mu}_{CP}$ reaches values of order 0.1, which
can be detected at a future muon collider \cite{cpkoldaetal}.

6.- In conclusion,
we have studied the effects of non-standard soft-breaking terms in
the MSSM, and found that they modify the chirality-changing
(LR) components of the squared sfermion mass matrices, which can
induce new sources of flavour violation.
Given present FCNC data, we can only estimate
the $A$ and $C$ parameters. To probe their strength,
we evaluate the decays $t \to c + h_i$, and find a B.R.
that may be detectable at LHC.
The C-terms also give the possibility to
explain the newly observed CP-violation phenomena as a SUSY
effect, and to measure a CP-violating higgs-lepton
coupling at a future muon collider.

{{\bf Acknowledgment.-} Discussions with G. Kane and M.A. Perez
are acknowledged. This work was supported by CONACYT and SNI (M\'exico).}

\newpage

\bigskip

{\bf TABLE CAPTION}

Table 1. B.R. of top FCNC decay $t \to c+h_i$.
Results are shown for $\tan\beta=2$, $m_{\tilde q}=300$ GeV,
$m_{\tilde g}=A^u=C^u=500$ GeV, and
the numbers in paranthesis correspond to 
$\tan\beta=10$.
\bigskip

\bigskip
\begin{center}
\begin{tabular}{||l|l|l| l|}
\hline
$m_A$ GeV & $B.R.(t \to c+h^0 )$  &  $B.R.(t \to c+H^0 )$  &
 $B.R.(t \to c+A^0 )$\\
\hline 
 100. &  $7.1   \times 10^{-4}$ ($4.8 \times 10^{-4}  $)  &
         $1.9   \times 10^{-5}$ ($1.1 \times 10^{-5}  $)  &
         $5.8   \times 10^{-4}$ ($3.8 \times 10^{-4}  $)  \\
\hline
 130. &  $7.0  \times 10^{-4}$ ($5.1  \times 10^{-4}$)  &
         $1.2  \times 10^{-6}$ ($1.7  \times 10^{-7}$)  &
         $3.9  \times 10^{-4}$ ($2.6  \times 10^{-4}$)  \\
\hline
 160. &  $6.8  \times 10^{-4}$  ($3.8 \times 10^{-4}$)  &
         $0$                   ($2.5  \times 10^{-5}$)  &
         $1.4 \times 10^{-4}$  ($9.6  \times 10^{-5}$)  \\
\hline
 190. &  $6.6  \times 10^{-4}$ ($3.3  \times 10^{-4}$)  &
         0                 (0)  &
         0                 (0)  \\
\hline
\end{tabular}
\end{center}

\bigskip

{\bf \, \, \, Table. 1}

\end{document}